\PassOptionsToPackage{numbers,sort&compress,square}{natbib}
\documentclass{iopjournal}
\usepackage{graphicx,float} % Required for inserting images

\usepackage[english]{babel}
\usepackage{natbib}
\usepackage{amsmath, amssymb, mathtools, physics}
\usepackage{hyperref}
\usepackage[nameinlink]{cleveref}
\Crefname{equation}{Eq.}{Eqs.}
\Crefname{figure}{Fig.}{Figs.}
\usepackage{enumitem}
\usepackage{multirow}
\usepackage{tikz}
\usetikzlibrary{external,cd}
\tikzsetexternalprefix{pdf/}
\usepackage{pgfplots}
\usepgfplotslibrary{colormaps} %colorbrewer

\usetikzlibrary{automata, arrows.meta, positioning, calc,shapes, pgfplots.groupplots}

\newcommand{\avg}[1]{\left\langle #1 \right\rangle}

\newcommand{\avgt}[1]{\langle #1|t \rangle}
\newcommand{\avgT}[1]{\langle #1|T \rangle}

\DeclareMathOperator{\artanh}{artanh}

\RenewDocumentCommand{\pdv}{ o m m }{\IfValueTF{#1}{\partial^{#1}_{#3} #2}{\partial_{#3} #2}}

\date{14.08.2026}

\begin{document}

\articletype{Paper} %    e.g. Paper, Letter, Topical Review...

\title{Generalizing the multidimensional thermodynamic uncertainty relation to combinations of arbitrary counting variables}

\author{Niklas Buschmann$^1$\orcid{0009-0005-6391-3645}, Udo Seifert$^1$\orcid{0000-0002-9271-6190} and Alexander M. Maier$^{1,*}$\orcid{0009-0001-5665-2716}}

\affil{$^1$II. Institut f\"ur Theoretische Physik, Universit\"at Stuttgart, 70550 Stuttgart, Germany}

\affil{$^*$Author to whom any correspondence should be addressed.}

\email{amaier@theo2.physik.uni-stuttgart.de}


\begin{abstract}
  Uncertainty relations provide lower bounds for otherwise hidden quantities of a partially accessible Markov network like the mean entropy production rate and the total dynamical activity. The thermodynamic uncertainty relation (TUR) is arguably the most prominent one and involves the precision of a fluctuating net current. One of its major generalizations is the multidimensional TUR (MTUR), which yields a tighter bound by using covariances of a set of observed net currents. We generalize this latter bound to time-dependently driven processes with arbitrary initial state and arbitrary measurement duration. Furthermore, this bound can be used with a set of fluctuating counting observables each of which can be time-antisymmetric, time-symmetric or time-asymmetric, i.e., a net current, a traffic or a flow. We can even allow for coarse-grained observations in which each counting observable could consist of multiple indiscernable observed transitions of the underlying system. Thus, we generalize the MTUR, extensions of the TUR for time-dependent processes based on one current, and an estimator based on one flux or on one traffic in a unifying way. We illustrate this general uncertainty relation with simple examples.

\end{abstract}

%================================================================================================================================[ SECTION ]===========%
\section{Introduction}

Stochastic thermodynamics is a framework for describing small nonequilibrium systems with thermodynamic quantities such as entropy production, work and heat that fluctuate along individual trajectories \cite{seki10,jarz11,peli21,shir23,seif25}. Since biophysical and chemical systems, like molecular motors and chemical reaction networks, are typically driven out of equilibrium, the entropy production quantifying irreversibility and energy dissipation of these systems is arguably one of the most important thermodynamic quantities. However, determining its rate for a given system is often impossible because contributing degrees of freedom are hidden.

The growing field of thermodynamic inference provides inference methods \cite{seif19} driven by the desire to infer hidden and incompletely observable thermodynamic quantities and other information on partially accessible physical systems like their topology \cite{maie25,zhao25}. These methods typically provide universal inequalities, or tradeoff relations, for a thermodynamic quantity based on observables. Prominent examples of such inequalities are speed limits describing the tradeoff between the time required to transform an initial distribution into a desired final one and the necessary dissipation to do this \cite{shir18,naka21,dech22,naga25,bao25}, and the thermodynamic uncertainty relation (TUR) using the precision of an observable fluctuating current given as the ratio of its squared mean and its variance to yield a lower bound on the mean entropy production rate (EPR) \cite{bara15,ging16}.
In recent years, a multitude of generalizations of the TUR have been found ranging from extensions to systems with time-dependent driving \cite{koyu20} and with a finite measurement duration \cite{piet17,horo17,dech18a,liu20,mani20,vanv20,otsu20,otsu22} to improvements based on correlations of observed currents \cite{dech21,ohga23} and on insights from optimal transport \cite{vanv23}. The multidimensional TUR (MTUR) makes use of covariances between discernible fluctuating currents \cite{dech18,more25} thus being a particularly important extension. Moreover, similar inequalities based on the precision of a counting observable such as a kinetic uncertainty relation (KUR) \cite{gara17,dite19,hiur21,shir23}, which provides a bound on the total dynamical activity, a clock uncertainty relation (CUR) \cite{prec25}, and hierarchies of tradeoff relations including the TUR \cite{kwon24,kwon26} have been derived.

Two more recent lines of research regarding inferring the EPR of a given system focus on lower bounds based on time-symmetric observations \cite{maes17,vo20,ziyi23,piet24} and on coarse-graining \cite{espo12}. In the case of coarse-grained observations involving lumped states, which are clusters of microscopic states being recorded as one mesostate, and blurred transitions, i.e., different transitions within an underlying Markov network that are observed as one transition, lower, though somewhat looser, bounds on the EPR can still be found \cite{gode22,erte24,piet24,haru24,igos25a,seif25a}. For these situations, EPR estimators for processes of finite duration and for time-dependent systems are still lacking to the best of our knowledge. Inspired by Refs. \cite{dech18,piet24}, here we will derive such lower bounds on the EPR based on general counting observables that can all be symmetric, antisymmetric or asymmetric in time. We go even one step further by deriving a generalized MTUR (GMTUR), which yields a lower bound on the EPR of a process making use of covariances between general counting observables accumulated over a finite time.

This paper is structured as follows. In Section \ref{sec:setup}, we describe the broad class of systems we consider. We will then derive a lower bound for the entropy production rate in Section \ref{sec:main}, first for the simpler case of a system with time-independent transition rates, and later on a generalized version for systems with time-dependent ones. In Section \ref{sec:1d}, we compare our result to extant ones for the case in which we can measure only one flow, one traffic or one current observable and its variance. In Section \ref{sec:2d}, we specialize our main result to all variants with two observables providing explicit expressions and some illustrations. We conclude the paper in Section \ref{sec:summary} and give an outlook on possible extensions.

%================================================================================================================================[ SECTION ]===========%
\section{Setup}\label{sec:setup}
%======================================================================================================================[ FIGURE ]
\begin{figure}[H]
    \centering
    % \tikzsetnextfilename{setup_figure}
    % \input{pdf/setup_figure.tikz}
    \includegraphics[scale=1]{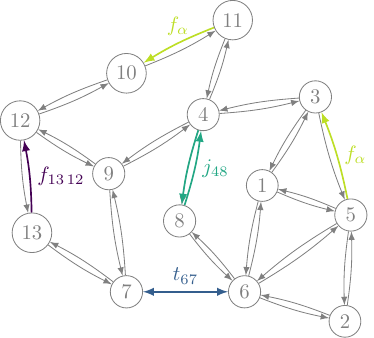}
    \caption{Partially observable Markov network with 13 discrete states. Gray transitions and states are hidden, whereas an observer may register the flow $f_{13\,12}$ (purple), the traffic $t_{67}$ (blue), the flows of the net current $j_{48}$ (green), the coarse-grained flow $f_\alpha=f_{53}+f_{11\,10}$ (yellow), or a combination of these.}
    \label{fig:setup}
\end{figure}
%==================================================================================================================[ FIGURE END ]

We consider continuous-time Markov jump processes on a set of discrete states $i,j,\dots$ with irreducible topology. The stochastic dynamics is characterized by transition rates $k_{ij}$ for transitions from state $i$ to state $j$. The rates $k_{ij}$ and $k_{ji}$ are both nonzero if a pair of transitions exists between states $i$ and $j$, which we call a link. Moreover, all transition rates are assumed to be time-independent unless indicated otherwise. The transition rates generate the evolution of the probability $p_i(t)$ to be in state $i$ at time $t$ through the master equation
\begin{align}\label{eq:master}
    \pdv{{p}_i(t)}{t}=\sum_{j}[p_j(t)k_{ji}-p_i(t)k_{ij}].
\end{align}

We assume that only some transitions or, more generally, sums of an unknown number thereof are observable. These sums are of one of three types. For one link between states $i$ and $j$, observations are (i) a flow $f_{ij}$ along transition $i\to j$ with invisible flow $f_{ji}$ along the reverse transition, (ii) a traffic $t_{ij} = f_{ij} + f_{ji}$ or (iii) a net current $j_{ij} = f_{ij} - f_{ji}$. In the general case, these observables will be called generalized rates and denoted by $r_\alpha\in\{f_\alpha,t_\alpha,j_\alpha\}$. Each discernible observation may consist of multiple indistinguishable contributions of the same type from different links. We label each distinct observation by an index $\alpha$, which we call the \emph{color} of the observable. We illustrate this in Fig. \ref{fig:setup} where an observer may be able to discern four colors, i.e., a current, a traffic and two flows, one of which is a sum of two flows that are not resolved separately.

During an experiment, the Markov network traces a trajectory $\gamma(t)$ of length $T$. Since we only record transitions contributing to generalized rates $r_\alpha(t)$, there are two meaningful averages of these fluctuating rates. First, the average at a given time $t$ over all possible trajectories for a fixed initial distribution is given by
\begin{align}
    \avgt{r_\alpha} =\sum_{ij \in \alpha}\avgt{r_{ij}}.
\end{align}
These averages are given by
\begin{align}
    \avgt{r_{ij}} &= \begin{cases}
        p_i(t)k_{ij} &\quad (\text{flow}), \\
        p_i(t)k_{ij}+p_j(t)k_{ji} &\quad (\text{traffic}), \\
        p_i(t)k_{ij}-p_j(t)k_{ji} &\quad (\text{current}).
    \end{cases}
\end{align}
Second, the time average of $\avgt{r_\alpha}$ within the interval $[0,T]$ will be denoted as
\begin{align}
    \avg{r_\alpha} \equiv \frac{1}{T}\int_0^T\avgt{r_\alpha}\mathrm{d}t,\label{time_integral}
\end{align}
with $T$ notationally suppressed throughout.
Similar to the mean rates, we determine their covariances
\begin{align}
    \Sigma_{\alpha\beta} = \avg{(r_\alpha-\avg{r_\alpha})(r_\beta-\avg{r_\beta})}
\end{align}
by averaging over trajectories and time $t$.
Moreover, the instantaneous entropy production rate $\sigma(t)$ of the system is given by \cite{peli21,shir23,seif25}
\begin{align}
    \sigma(t) \equiv \sum_{ij}\sigma_{ij}(t) =\sum_{ij}p_i(t)k_{ij}\ln\frac{p_i(t)k_{ij}}{p_j(t)k_{ji}}=\sum_{ij} \avgt{j_{ij}}\artanh\left(\frac{\avgt{j_{ij}}}{\avgt{t_{ij}}}\right). \label{def:epr}
\end{align}

%================================================================================================================================[ SECTION ]===========%
\section{Derivation of the GMTUR}\label{sec:main}
%================================================================================================================================[ SUBSECTION ]--------%
\subsection{Multivariate Cram\'er-Rao bound and auxiliary dynamics}
Let $\mathbb{P}[\gamma(t)]$ denote the path weight of the trajectory $\gamma(t)$, so that $\avg{.}$ is the average with respect to these path weights. Additionally, let $\theta$ be a parameter controlling the strength of a perturbation that modifies the rates $k_{ij}$ to $k_{ij}^{\theta}$ without changing the initial probabilities $p_i(0)$. The path weight $\mathbb{P}^{\theta}[\gamma(t)]$ of such an auxiliary process reads
\begin{align}\label{pathweight}
    \mathbb{P}^{\theta}[\gamma(t)] = p_{\gamma(0)}(0)e^{-\sum_{ij} k_{ij}^{\theta} \tau_i[\gamma(t)]}\prod_{ij}(k_{ij}^{\theta})^{n_{ij}[\gamma(t)]}
\end{align}
where $\tau_i[\gamma(t)]$ is the time spent in state $i$ and $n_{ij}[\gamma(t)]$ is the number of jumps $i\to j$ along $\gamma(t)$ up to time $T$. Corresponding averages are denoted by $\langle .\rangle_{\theta}$.

The multivariate Cram\'er-Rao bound \cite{rao45,cram99} provides a lower bound on the Fisher information in terms of the derivatives of mean rates $\partial_\theta \langle\boldsymbol{r}\rangle_{\theta}$ and their inverse covariance matrix $\boldsymbol{\Sigma}^{-1}(\theta)$ as
\begin{align}
     I(\theta)=\avg{\left(\pdv{ \ln\mathbb{P}^\theta[\gamma(t)]}{\theta}\right)^2 }_\theta\geq \sum_{\alpha\beta}[\pdv{\avg{r_\alpha}_\theta }{\theta}] \Sigma^{-1}_{\alpha\beta}(\theta) [\pdv{\avg{r_\beta}_\theta}{\theta}].\label{cramer-rao}
\end{align}
In the following, we first evaluate the Fisher information $I(\theta)$ for a perturbation that (i) relates the Fisher information to the entropy production rate $\sigma$ and (ii) yields derivatives $\partial_\theta \avg{r_\alpha}_\theta$ that are proportional to the mean rates $\avg{r_\alpha}$.
%================================================================================================================================[ SUBSECTION ]--------%
\subsection{Evaluating the Fisher information}
The left-hand side of inequality \eqref{cramer-rao} admits the form
\begin{align}
    I(\theta) &= \avg{\frac{1}{ \mathbb{P}^\theta[\gamma(t)]}\pdv[2]{\mathbb{P}^\theta[\gamma(t)]}{\theta}}_\theta - \avg{\pdv[2]{\ln \mathbb{P}^\theta[\gamma(t)]}{\theta}}_\theta
    = \avg{\pdv[2]{}{\theta}\sum_{ij} k_{ij}^\theta \tau_i[\gamma(t)] - n_{ij}[\gamma(t)]\ln k_{ij}^\theta}_\theta.\label{pre_dechant}
\end{align}
Note that the first average in the middle of Eq. \eqref{pre_dechant} vanishes since path weights sum to unity independent of $\theta$.
The average time spent cumulatively in state $i$ along all trajectories $\gamma(t)$ up to their final time $t=T$ simplifies to $\avg{\tau_i[\gamma(t)]}_\theta = \int_0^T p_i^\theta(t)\mathrm{d}t$. Here, $p_i^\theta(t)$ is the solution of the master equation \eqref{eq:master} for the auxiliary dynamics. Similarly, the average number of jumps $i\to j$ is given by $\avg{n_{ij}[\gamma(t)]}_\theta = \int_0^T p_i^\theta(t)k_{ij}^\theta\mathrm{d}t$, resulting in the Fisher information
\begin{align}
    I(\theta) = \int_0^T\sum_{ij} p_i^\theta(t)\left[\pdv[2]{k_{ij}^\theta}{\theta} - k_{ij}^\theta\pdv[2]{\ln k_{ij}^\theta}{\theta}\right]\mathrm{d}t,\label{res_dechant}
\end{align}
which can, e.g., also be found in Ref. \cite{liu20}.

We now use the perturbation
\begin{align}\label{perturbation}
    k_{ij}^\theta = k_{ij}e^{\theta[x_{ij}(t)(1-z_{ij})+z_{ij}]},
\end{align} where $z_{ij}=z_{ji}$ is arbitrary and $x_{ij}(t)\equiv \avgt{j_{ij}}/\avgt{t_{ij}}\in(-1,1)$ is the antisymmetric ratio of current and traffic along the link between $i$ and $j$. For this perturbation, the Fisher information becomes
\begin{align}\label{after_dechant}
    I(\theta) &= \int_0^T\sum_{ij} p_i^\theta(t) k_{ij}^\theta[x_{ij}(t)(1-z_{ij}) + z_{ij}]^2\mathrm{d}t.
\end{align}
Using the relation $p_i(t)k_{ij}=\avgt{t_{ij}}[1+x_{ij}(t)]/2$, we find
\begin{align}
    I(0)&= \int_0^T\sum_{i<j} \avgt{t_{ij}} [x_{ij}^2(t)(1-z_{ij}^2)+z_{ij}^2]\mathrm{d}t
\end{align}
where we omit the vanishing antisymmetric terms.
When setting $z_{ij}=0$, this expression of the Fisher information is a lower bound on half of the EPR $\sigma$ since
\begin{align}
    I(0)\big|_{z_{ij}=0}&= \int_0^T \sum_{i<j} \avgt{t_{ij}} x_{ij}^2(t)\mathrm{d}t \ \leq \ \int_0^T\sum_{i<j} \avgt{t_{ij}} x_{ij}(t)\artanh(x_{ij}(t))\mathrm{d}t = \frac{1}{2} \int_0^T\sum_{ij} \sigma_{ij}(t) \mathrm{d}t. \label{epr_fisher}
\end{align}
Choosing $\ z_{ij} = z_\alpha $ for all links $ij\in \alpha$ where traffic or flow is observed and $z_{ij}=0$ otherwise, yields
\begin{align}
    I(0) \leq \int_0^T\biggl\{\frac{\sigma(t)}{2}&+\sum_\alpha\sum_{ij\in \alpha} \avgt{t_{ij}}\left[x_{ij}^2(t)(1-z_\alpha^2)+z_\alpha^2-x_{ij}(t)\artanh\left(x_{ij}(t)\right)\right] \notag\\
    &+\sum_\beta\sum_{ij\in \beta} \frac{2\avgt{f_{ij}}}{1+x_{ij}(t)}\left[x_{ij}^2(t)(1-z_\beta^2)+z_\beta^2-x_{ij}(t)\artanh(x_{ij}(t))\right]\biggr\}\mathrm{d}t.\label{pre_avg}
\end{align}
Here, the second line arises by expressing the traffic in terms of the flow.

The bound \eqref{pre_avg} attains a simpler form by denoting the averages of $x_{ij}(t)$ and $x_{ij}^2(t)$ as
\begin{align}
x_\alpha \equiv \frac{\int_0^T\sum_{ij\in\alpha} \avgt{f_{ij}} x_{ij}(t)\mathrm{d}t}{\int_0^T\sum_{ij\in\alpha} \avgt{f_{ij}}\mathrm{d}t}\in (-1,1) \quad \text{and} \quad y_\alpha \equiv \frac{\int_0^T\sum_{ij\in\alpha} \avgt{t_{ij}} x_{ij}^2(t)\mathrm{d}t}{\int_0^T\sum_{ij\in\alpha} \avgt{t_{ij}}\mathrm{d}t}\in(0,1) \label{averages}
\end{align}
where the denominators are equal to $T\avg{r_\alpha}$ as defined in \eqref{time_integral}. In addition to these definitions, we now use that the functions $\sqrt{y}\artanh(\sqrt{y})-y(1-z^2)-z^2$ and $[x\artanh(x)-x^2(1-z^2)-z^2]/(1+x)$ are convex. Then, applying Jensen's inequality to \eqref{pre_avg}, we arrive at
\begin{align}
    \frac{I(0)}{T} \leq \frac{\sigma}{2}&+\sum_\alpha \avg{t_\alpha} \left[y_\alpha(1-z_\alpha^2)+z_\alpha^2-\sqrt{y_\alpha}\artanh(\sqrt{y_\alpha})\right]\notag\\
    &+\sum_\beta \frac{2\avg{f_\beta}}{1+x_\beta} \left[x_\beta^2(1-z_\beta^2)+z_\beta^2-x_\beta\artanh(x_\beta)\right] \label{bound_fisher}
\end{align}
where $\sigma=\int_0^T\sigma(t)\mathrm{d}t/T$ denotes the mean entropy production rate along the trajectories $\gamma(t)$ of length $T$.

%================================================================================================================================[ SUBSECTION ]--------%
\subsection{Mean rates in the auxiliary dynamics} \label{subsec:meanrates}
We now turn to the right-hand side of inequality \eqref{cramer-rao}, which we evaluate in the auxiliary dynamics with
\begin{align}
    p_i^\theta(t) = p_i(t) + \theta t\partial_t p_i(t)+O(\theta^2) \quad \text{and} \quad k_{ij}^\theta = k_{ij} + \theta k_{ij}[x_{ij}(t)(1-z_{ij})+z_{ij}] +O(\theta^2).
\end{align}
For this dynamics, the master equation \eqref{eq:master} holds up to first order in $\theta$ since inserting $p_i^\theta(t)$ and $ k_{ij}^\theta$ there yields
\begin{align}
    \partial_t p_i^\theta=\partial_t p_i+\theta \partial_t p_i+\theta t\partial_t^2 p_i + O(\theta^2)= \sum_{j}(p_j+\theta t \partial_tp_j)k_{ji}^\theta- (p_i+\theta t \partial_tp_i)k_{ij}^\theta +O(\theta^2),\label{dot_p}
\end{align}
where we omit the time arguments for better readability.
Therefore, the derivative of a mean rate \eqref{time_integral} with respect to $\theta$,
\begin{align}
    \pdv{\avg{r_\alpha}_\theta}{\theta} &= \frac{1}{T}\int_0^T\sum_{ij\in \alpha}\pdv{\avgt{r_{ij}}_\theta}{\theta}\mathrm{d}t,
\end{align}
admits the form
\begin{align}
    \left.\pdv{\avg{f_\alpha}}{\theta}_\theta\right|_{\theta=0} &= \frac{1}{T}\int_0^T\sum_{ij\in \alpha} \avgt{f_{ij}}[x_{ij}(t)(1-z_\alpha) + z_\alpha]+tk_{ij}\partial_t p_i(t)\mathrm{d}t, \label{rates_1} \\
    \left.\pdv{\avg{t_\alpha}}{\theta}_\theta\right|_{\theta=0} &= \frac{1}{T}\int_0^T\sum_{ij\in \alpha} \avgt{t_{ij}}[x_{ij}^2(t)(1-z_\alpha) + z_\alpha]+tk_{ij}\partial_t p_i(t)+tk_{ji}\partial_t p_j(t)\mathrm{d}t, \label{rates_2}\\
    \left.\pdv{\avg{j_\alpha}}{\theta}_\theta\right|_{\theta=0} &= \frac{1}{T}\int_0^T\sum_{ij\in \alpha} \avgt{j_{ij}}+tk_{ij}\partial_t p_i(t)-tk_{ji}\partial_t p_j(t)\mathrm{d}t\label{rates_3}
\end{align}
for flow, traffic and current, respectively.
Integrating Equations \eqref{rates_1} to \eqref{rates_3} by parts and using the coefficients defined in \eqref{averages}, we arrive at
\begin{align}
    \left.\pdv{\avg{r_\alpha}}{\theta}_\theta\right|_{\theta=0} &= \begin{cases}
         \avgT{f_\alpha}+[x_\alpha(1-z_\alpha)+z_\alpha-1] \avg{f_\alpha} , \\
         \avgT{t_\alpha}+[y_\alpha(1-z_\alpha)+z_\alpha-1] \avg{t_\alpha}, \\
        \avgT{j_\alpha} .\label{full_rates}
    \end{cases}
\end{align}
Moreover, in the long-time limit, the final and mean rates approach their stationary value, and thus
\begin{align}
    \lim_{T\to \infty}\left.\pdv{\avg{r_\alpha}}{\theta}_\theta\right|_{\theta=0} &= \begin{cases}
         [x_\alpha(1-z_\alpha)+z_\alpha] \avg{f_\alpha} &\quad (\text{flow}), \\
         [y_\alpha(1-z_\alpha)+z_\alpha] \avg{t_\alpha}&\quad (\text{traffic}), \\
        \avg{j_\alpha} &\quad (\text{current}).
    \end{cases}\label{rates}
\end{align}

%================================================================================================================================[ SUBSECTION ]--------%
\subsection{Main result}
We now insert the upper bound \eqref{bound_fisher} of the Fisher information, the derivative of the mean rates \eqref{rates} and $\boldsymbol{\Sigma}^{-1} \equiv \boldsymbol{\Sigma}^{-1}(0)$ into the Cram\'er-Rao bound \eqref{cramer-rao}. The resulting inequality yields our first main result, the lower bound on the EPR
\begin{align}
    \frac{\sigma}{2} \geq \min_{\mathbf{x},\mathbf{y}}\max_{\mathbf{z}}&\ \frac{1}{T}\sum_{\alpha\beta}\Sigma_{\alpha\beta}^{-1} \left.\left[\pdv{\avg{r_\alpha}}{\theta}_\theta\right]\left[\pdv{\avg{r_\beta}}{\theta}_\theta\right]\right|_{\theta=0} \notag\\
    &+ \sum_{\alpha} \avg{t_\alpha}\left[\sqrt{y_\alpha}\artanh(\sqrt{y_\alpha})-y_\alpha-z_\alpha^2(1-y_\alpha)\right]\notag\\
    &+ \sum_{\beta} \frac{2\avg{f_\beta}}{1+x_\beta}\left[x_\beta\artanh(x_\beta)-x_\beta^2-z_\beta^2(1-x_\beta^2)\right],\label{main}
\end{align}
which is thus a thermodynamically consistent entropy estimator.
Since in this bound the $x_\alpha\in(-1,1)$ and the $y_\alpha\in(0,1)$ are unknown, we have to minimize with respect to them. In contrast, the bound is valid for arbitrary $z_\alpha$, so we maximize with respect to the latter to get the strongest possible bound on $\sigma$.

%================================================================================================================================[ SUBSECTION ]--------%
\subsection{Simple bound}
Before we transform inequality \eqref{main} by analytically performing the maximization, we consider a particular simplified version. In the long-time limit $T\to \infty$ and for the suboptimal choice $z_\alpha=1$, inequality \eqref{main} reduces to
\begin{align}
    \frac{\sigma}{2}+\avg{t}+c \avg{f} &\geq \frac{1}{T}\sum_{\alpha\beta}\Sigma_{\alpha\beta}^{-1} \avg{r_\alpha}\avg{r_\beta} \label{simplified}
\end{align}
where $\avg{t} \equiv \sum_\alpha \avg{t_\alpha}$ and $\avg{f} \equiv \sum_\beta \avg{f_\beta}$ denote the total observed traffic and flow rates, respectively, and $c=-2\min_x (x\artanh(x)-1)/(1+x)\simeq 2.93$.
The bound \eqref{simplified} can thus be seen as an extension of the multidimensional TUR \cite{dech18}
\begin{align}
    \frac{\sigma}{2} &\geq \frac{1}{T}\sum_{\alpha\beta}\Sigma_{\alpha\beta}^{-1} \avg{j_\alpha}\avg{j_\beta},\label{MTUR}
\end{align}
which is restricted to current observables.

%================================================================================================================================[ SUBSECTION ]--------%
\subsection{Maximization over z}
The maximization over $z$ in inequality \eqref{main} is analytically possible since the equation is quadratic in $z$. For the sake of readability, we here restrict ourselves to the long-time limit $T\to\infty$, i.e., to the steady state, where we can use the simpler expressions \eqref{rates}. We substitute $u_\alpha^f\equiv x_\alpha+z_\alpha(1-x_\alpha)$ for the flow and $u_\alpha^t\equiv y_\alpha+z_\alpha(1-y_\alpha)$ for the traffic into Equations \eqref{rates} and \eqref{main}. Next, we define the coefficients
\begin{align}
    \kappa_\alpha^f&\equiv\frac{2\avg{f_\alpha}}{1-x_\alpha}, \ \lambda_\alpha^f\equiv\frac{\avg{f_\alpha}}{T}\sum_\mu\Sigma^{-1}_{\alpha\mu}\avg{ j_\mu} +\frac{2\avg{f_\alpha} x_\alpha}{1-x_\alpha},\  \rho_\alpha^f\equiv 2\avg{f_\alpha}\left[\frac{x_\alpha\artanh(x_\alpha)}{1+x_\alpha}-\frac{2x_\alpha^2}{1-x_\alpha^2}\right]\label{co2}
\end{align}
and
\begin{align}
    \kappa_\alpha^t&\equiv\frac{\avg{t_\alpha}}{1-y_\alpha},\  \lambda_\alpha^t\equiv\frac{\avg{t_\alpha}}{T}\sum_\mu\Sigma^{-1}_{\alpha\mu}\avg{j_\mu} +\frac{\avg{t_\alpha} y_\alpha}{1-y_\alpha},\  \rho_\alpha^t\equiv\avg{t_\alpha}\left[\sqrt{y_\alpha}\artanh(\sqrt{y_\alpha})-\frac{y_\alpha}{1-y_\alpha}\right] \label{co1}
\end{align}
corresponding in the quadratic equation \eqref{main} to the powers of $u_\alpha$ for the flow and the traffic, respectively.
Additionally, we denote the term stemming solely from the current observables in \eqref{main} as $\omega\equiv 2\sum_{\alpha\beta}\Sigma^{-1}_{\alpha\beta}\avg{j_\alpha} \avg{j_\beta}/T$ and define the matrix $\boldsymbol{M}$ with elements
\begin{align}
    M_{\alpha\beta} \equiv \delta_{\alpha\beta}\kappa_\alpha-\frac{1}{T}\Sigma_{\alpha\beta}^{-1}\avg{r_\alpha}\avg{r_\beta}
\end{align}
for indices $\alpha,\beta$ corresponding to traffic or flow using the Kronecker delta $\delta_{\alpha\beta}$. With these definitions, our first main result \eqref{main} attains the simple form
\begin{align}
    \frac{\sigma}{2} &\geq \frac{\omega}{2}+\min_{\mathbf{x},\mathbf{y}}\max_{\mathbf{u}}\left[ -\sum_{\alpha\beta}M_{\alpha\beta}u_\alpha u_\beta +2\sum_\alpha\lambda_\alpha u_\alpha + \sum_\alpha\rho_\alpha\right] \notag\\
    &= \frac{\omega}{2} +\min_{\mathbf{x},\mathbf{y}:M\succeq0} \left[\sum_{\alpha\beta}M^{-1}_{\alpha\beta}\lambda_\alpha\lambda_\beta  + \sum_\alpha \rho_\alpha\right],\label{minimized}
\end{align}
with  $x_\alpha\in(-1,1)$ and $y_\alpha\in(0,1)$ in the limit $T\to\infty$. Using the second line, we only need to perform a minimization over all $x_\alpha$ and $y_\alpha$ under the constraint that the matrix $\boldsymbol{M}$ must be positive semi-definite. A complementary derivation via large deviation theory yields the same result for arbitrary sets of observable traffics and flows, which we discuss in Appendix \ref{app:complderiv}. The bound \eqref{minimized} can be saturated for a specific class of networks, which we discuss in Appendix \ref{app:saturation}.

%================================================================================================================================[ SUBSECTION ]--------%
\subsection{Time-dependently driven systems}
So far, we have assumed the transition rates $k_{ij}$ to be time-independent. Extending our main result \eqref{main} to time-dependently driven systems requires only minor modifications, which we will discuss now.
We assume that the transition rates  $k_{ij}(\eta)$ become time-dependent through a protocol $\eta(vt)$ with speed parameter $v$. Then, the path weight \eqref{pathweight} must be modified to
\begin{align}
    \mathbb{P}^\theta[\gamma(t)] = p_{\gamma(0)}\exp\left(-\int_0^T\Gamma_{\gamma(t)}^\theta(t)\mathrm{d}t\right)\prod_{t_l} k_{\gamma^-(t_{l}) \gamma^+(t_{l})}^\theta(t_l)
\end{align}
with the escape rate $\Gamma_{i}^\theta(t) \equiv \sum_j k_{ij}^\theta(t)$ for state $i = \gamma(t)$ and the times $\{t_l\}$ at which a jump from state $\gamma^-(t_l)$ to state $\gamma^+(t_l)$ occurs, which are the left-sided and right-sided limit of $\gamma(t)$ at $t_l$.
Since we choose the perturbed transition rates to be of the form $k^\theta_{ij}(t)=e^{\theta f(x,z,t)}k_{ij}(t)$, we have $\partial_\theta^2\ln k^\theta_{ij}(t) = 0$ and find the Fisher information
\begin{align}
    I(\theta) &= \avg{\int_0^T\pdv[2]{}{\theta}\Gamma_{\gamma(t)}^\theta(t)\mathrm{d}t}_\theta
    = \int_0^T \sum_{i}p_i^\theta(t) \pdv[2]{}{\theta}\Gamma_i^\theta(t)\mathrm{d}t= \int_0^T \sum_{ij} p_i^\theta(t) \pdv[2]{}{\theta}k_{ij}^\theta(t)\mathrm{d}t.
\end{align}
Thus, for our choice of rates in the auxiliary dynamics, Equation \eqref{res_dechant} also applies to systems with time-dependent transition rates.

From there on, we can use the derivation for time-independent transition rates until we need to determine the $\theta$ derivative of mean rates like in Section \ref{subsec:meanrates}, where the time-dependence of the transition rates must be taken into account. In particular, we have to use
\begin{align}
  p_i^\theta(t) &= p_i(t) + \theta\left[t\partial_t p_i(t) - v\partial_v p_i(t)\right] + O(\theta^2)
\end{align}
for the perturbed probability such that it again solves the master equation \eqref{eq:master} with the transition rates $k^\theta_{ij}(t)$ of the auxiliary dynamics up to first order in $\theta$. Using this modified probability for the auxiliary system, the integrands in Equations \eqref{rates_1} to \eqref{rates_3} contain an extra term $-v [\partial_v p_i(t) ]k_{ij}(t)$ per contributing flow $p_i(t)k_{ij}(t)$. Writing the transition rates as $k_{ij}(\eta(vt))$, these integrals can be evaluated using the relation
\begin{align}
    \int_0^T tk_{ij}(\eta)\pdv{p_i(t)}{t} \mathrm{d}t &= [tp_i(t)k_{ij}(\eta)]_0^T -\int_0^T p_i(t)\left[k_{ij}(\eta)+t\pdv{k_{ij}(\eta)}{t}\right] \mathrm{d}t \notag \\
    &= Tp_i(T)k_{ij}(\eta(vT)) -\int_0^T p_i(t)k_{ij}(\eta)\mathrm{d}t \notag \\ &\qquad - v\pdv{}{v}\int_0^T p_i(t)k_{ij}(\eta)\mathrm{d}t + v\int_0^T \left[\pdv{}{v}p_i(t)\right]k_{ij}(\eta)\mathrm{d}t .
\end{align}
Therefore, the derivatives of the generalized rates, i.e., the analogs of Equation \eqref{full_rates}, are
\begin{align}
    \left.\pdv{\avg{r_\alpha}}{\theta}_\theta\right|_{\theta=0} &= \begin{cases}
         \avgT{f_\alpha}+[x_\alpha(1-z_\alpha)+z_\alpha-1-v\partial_v] \avg{f_\alpha} &\quad (\text{flow}), \\
         \avgT{t_\alpha}+[y_\alpha(1-z_\alpha)+z_\alpha-1-v\partial_v]\avg{t_\alpha}&\quad (\text{traffic}), \\
        \avgT{j_\alpha}-v\partial_v\avg{j_\alpha} &\quad (\text{current}).
    \end{cases} \label{modifiedrates}
\end{align}
These expressions highlight that time-dependently driven systems do not reach a steady state, i.e., the average $\avgT{ r_\alpha}$ does not converge in the long-time limit.

Inserting the quantities \eqref{modifiedrates} and using the appropriate mean rates and their covariances in our first main result \eqref{main} constitutes our second main result. Since the resulting expression differs from the first one only due to the modified $\theta$ derivative of the rates \eqref{modifiedrates}, we omit repeating it. Moreover, for a single net current $j_{ij}(t)$, the expression in Equation \eqref{modifiedrates} is equivalent to $\avg{j_{ij}} + T\partial_T \avg{j_{ij}} - v\partial_v \avg{j_{ij}}$. Thus, for a time-dependently driven system with arbitrary initial state and only one observable current, our time-dependent result specializes to the expression derived in Ref. \cite{koyu20}.

%================================================================================================================================[ SECTION ]===========%
\section{One color}\label{sec:1d}
In the special case of observing a single current in the steady state, the bound \eqref{main} yields the TUR, while, for a single traffic or flow observable, it is algebraically equivalent to the bound derived by Pietzonka and Coghi \cite{piet24}. Specifically, with the Fano factor
\begin{align}
    F\equiv T \Sigma/\avg{r} \label{fano}
\end{align}
and a single current, bound \eqref{minimized} becomes the TUR \cite{bara15,ging16}
\begin{align}
    \frac{\sigma}{\avg{j}} &\geq \frac{2}{F} = \frac{2\avg{j}}{T\Sigma}.\label{1d_tur}
\end{align}
For a single traffic observable, we find
\begin{align}
    \frac{\sigma}{\avg{t}} &\geq \min_{y\in(1-F,1)}\,2\sqrt{y}\artanh(\sqrt{y})+\frac{2y(1-F)}{F-1+y} \geq \frac{2}{F}-2.\label{1d_traffic}
\end{align}
Moreover, for a single flow observable, the bound \eqref{minimized} becomes
\begin{align}
    \frac{\sigma}{\avg{f}} &\geq \min_{x\in(1-2F,1)}\,\frac{1}{1+x}\left(4x\artanh(x)+\frac{8x^2(1-F)}{2F-1+x}\right) \geq \frac{2}{F}-2c \label{1d_flow}
\end{align}
where $2c\simeq 5.86$. Some algebra shows that the bounds \eqref{1d_traffic} and \eqref{1d_flow} are indeed equivalent to the ones in \cite{piet24}, as we outline in Appendix \ref{sec:equivalence}. Furthermore, the second inequality in \eqref{1d_traffic} and in \eqref{1d_flow} corresponds to the simple bound \eqref{simplified}.
%======================================================================================================================[ FIGURE ]
\begin{figure}
    \centering
    % \tikzsetnextfilename{bound_1d}
    % \input{pdf/bound_1d.tikz}
    \includegraphics[scale=1]{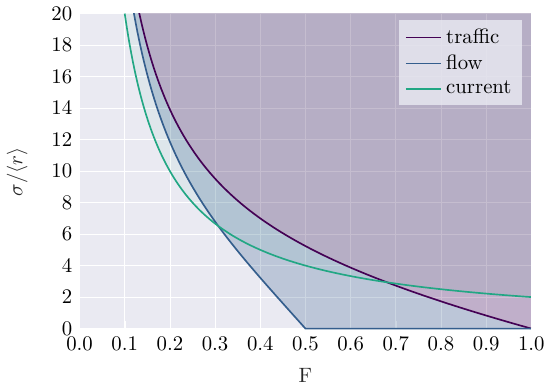}
    \caption{Comparison of the lower bounds \eqref{1d_tur} to \eqref{1d_flow} for one observed color per rate $r\in\{t,f,j\}$. For the traffic $t$ (purple), the flow $f$ (blue) and the current $j$ (green), the respective shaded region indicates permissible values of the steady-state entropy production rate $\sigma$ in units of the corresponding mean rate $\avg{r}$ as a function of the Fano factor $F$ as defined in \eqref{fano}.}
    \label{fig:1d}
\end{figure}
%==================================================================================================================[ FIGURE END ]

We illustrate the bound \eqref{minimized} for the three types of rates in Figure \ref{fig:1d}. In contrast to the TUR, a precision of $F\geq 1/2$ for the flow and $F\geq1$ for the traffic can be realized in equilibrium, e.g., by a two-state system. Moreover, for a Fano factor $F\lesssim 0.68$ for the traffic and $F\lesssim 0.31$ for the flow, the minimum entropy production required is higher than the one required for a current with the same level of precision.

%================================================================================================================================[ SECTION ]===========%
\section{Two colors}\label{sec:2d}
We now assume that an observer is able to register rates of two different colors for which there are several cases. First, for two currents, the bound \eqref{main} becomes the two-dimensional MTUR \cite{dech18}
\begin{align}
    \frac{\sigma}{2}\geq \frac{1}{T}\begin{pmatrix}
        \avg{j_1} \\\avg{j_2}
    \end{pmatrix}^\intercal\begin{pmatrix}
        \Sigma^{-1}_{11} & \Sigma^{-1}_{12} \\
        \Sigma^{-1}_{12} & \Sigma^{-1}_{22}
    \end{pmatrix}\begin{pmatrix}
        \avg{j_1} \\\avg{j_2}
    \end{pmatrix}.
\end{align}
Second, for two flows, we substitute $w_i \equiv 2x_i/(1-x_i)$ into inequality \eqref{minimized} and arrive at
\begin{align}
    \frac{\sigma}{2}\geq \min_{w_1,w_2:M\succeq0}&\begin{pmatrix}
        w_1\\ w_2
    \end{pmatrix}^\intercal\begin{pmatrix}
        \frac{w_1+2}{\avg{f_1}} - \frac{1}{T}\Sigma^{-1}_{11}& -\frac{1}{T}\Sigma^{-1}_{12} \\
        -\frac{1}{T}\Sigma^{-1}_{12} & \frac{w_2+2}{\avg{f_2}} - \frac{1}{T}\Sigma^{-1}_{22}
    \end{pmatrix}^{-1}\begin{pmatrix}
        w_1\\ w_2
    \end{pmatrix} \notag\\
    &+\frac{\avg{f_1}w_1}{1+w_1}\left[\artanh\left(\frac{w_1}{2+w_1}\right)-w_1\right]+\frac{\avg{f_2}w_2}{1+w_2}\left[\artanh\left(\frac{w_2}{2+w_2}\right)-w_2\right].\label{2d_flow}
\end{align}
Here, the minimization over $w_{1,2}$ is constrained by the requirement that the matrix in the first line must remain positive semi-definite.
Third, for two traffic observables and with the substitution $w_i \equiv y_i/(1 -y_i)$, bound \eqref{minimized} becomes
\begin{align}
    \frac{\sigma}{2}\geq \min_{w_1,w_2:M\succeq0}&\begin{pmatrix}
        w_1\\ w_2
    \end{pmatrix}^\intercal\begin{pmatrix}
        \frac{w_1+1}{\avg{t_1}} - \frac{1}{T}\Sigma^{-1}_{11}& -\frac{1}{T}\Sigma^{-1}_{12} \\
        -\frac{1}{T}\Sigma^{-1}_{12} & \frac{w_2+1}{\avg{t_2}} - \frac{1}{T}\Sigma^{-1}_{22}
    \end{pmatrix}^{-1}\begin{pmatrix}
        w_1\\ w_2
    \end{pmatrix} \notag\\
    &+\avg{t_1}\left[\sqrt{\frac{w_1}{1+w_1}}\artanh\left(\sqrt{\frac{w_1}{1+w_1}}\right)-w_1\right] \notag \\ &+\avg{t_2}\left[\sqrt{\frac{w_2}{1+w_2}}\artanh\left(\sqrt{\frac{w_2}{1+w_2}}\right)-w_2\right].\label{2d_traffic}
\end{align}
Fourth, for a mixture of a traffic and a flow, we find
\begin{align}
    \frac{\sigma}{2}\geq \min_{w_1,w_2:M\succeq0}&\begin{pmatrix}
        w_1\\ w_2
    \end{pmatrix}^\intercal\begin{pmatrix}
        \frac{w_1+1}{\avg{t}}-\frac{1}{T}\Sigma^{-1}_{tt} & -\frac{1}{T}\Sigma^{-1}_{tf} \\
        -\frac{1}{T}\Sigma^{-1}_{tf} & \frac{w_2+2}{\avg{f}}-\frac{1}{T}\Sigma^{-1}_{ff}
    \end{pmatrix}^{-1}\begin{pmatrix}
        w_1\\ w_2
    \end{pmatrix} \notag\\
    &+\avg{t}\left[\sqrt{\frac{w_1}{1+w_1}}\artanh\left(\sqrt{\frac{w_1}{1+w_1}}\right)-w_1\right]+\frac{\avg{f}w_2}{1+w_2}\left[\artanh\left(\frac{w_2}{2+w_2}\right)-w_2\right].\label{2d_traffic_flow}
\end{align}
Fifth, a mixture of a current and a traffic yields
\begin{align}
    \frac{\sigma}{2}\geq \min_{y} &\frac{\Sigma_{jj}^{-1}}{T}\avg{j}^2+\frac{1}{1-y}\frac{\left[\avg{j}\avg{t}(1-y) \Sigma^{-1}_{jt}+T\avg{t} y\right]^2}{T\left[T\avg{t}-(1-y)\avg{t}^2\Sigma_{tt}^{-1}\right]}+\avg{t}\left[ \sqrt{y_\alpha}\artanh(\sqrt{y})-\frac{y}{1-y}\right]
\end{align}
with $y\in(1-T/\avg{t} \Sigma_{tt}^{-1},1)$, and, finally, a mixture of a current and a flow leads to
\begin{align}
    \frac{\sigma}{2}\geq \min_{x} &\frac{\Sigma_{jj}^{-1}}{T}\avg{j}^2+\frac{1}{1-x}\frac{\left[\avg{j} \avg{f}(1-x) \Sigma^{-1}_{jf}+2T\avg{f} x\right]^2}{T\left[2T\avg{f} -\avg{f}^2(1-x)\Sigma_{ff}^{-1}\right]}+2\avg{f} \left[\frac{x\artanh(x)}{1+x}-\frac{2x^2}{1-x^2}\right]
\end{align}
with $x\in(1-2T/\avg{f} \Sigma_{ff}^{-1},1)$.

\begin{figure}
    \centering
    % \tikzsetnextfilename{bound_2d_alt}
    % \input{pdf/bound_2d_alt.tikz}
    \includegraphics[scale=1]{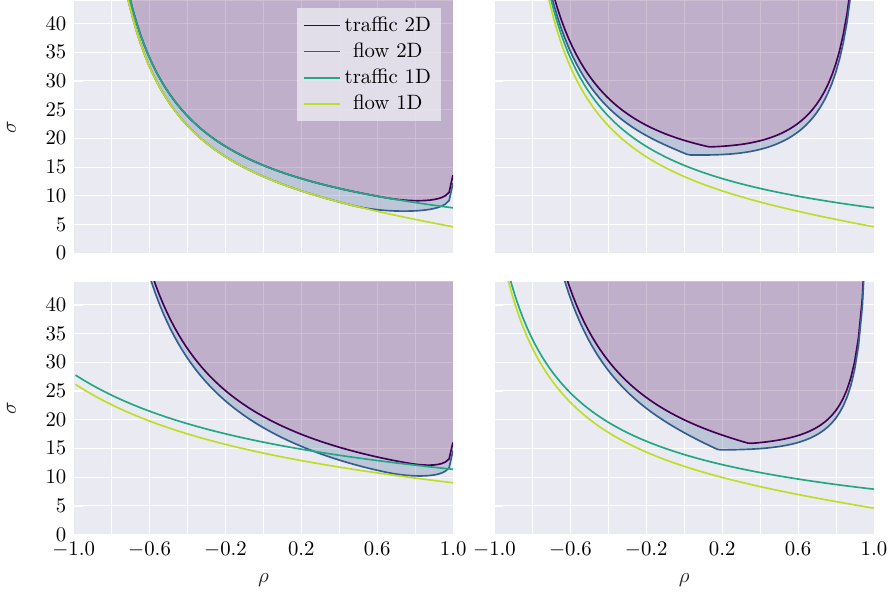}
        \caption{Comparison of bounds on the steady-state entropy production rate $\sigma$ based on one (1D) or two (2D) colors of traffic or flow. Bound \eqref{main} depends on five variables in the 2D case, of which we fix the mean rate $\avg{r_\alpha}$ and the standard deviation $\sigma_\alpha$ of the fluctuating traffic or flow of color $\alpha\in\{1,2\}$ and vary the correlation $\rho\equiv\sigma_1\sigma_2/\Sigma_{12}$ between $r_1$ and $r_2$ in each of the graphs. The fixed values $(\avg{r_1},\avg{r_2},\sigma_1,\sigma_2)$ are $(0.5,0.5,0.3,0.3)$ in the upper left panel, $(0.2,0.8,0.3,0.3)$ in the upper right panel, $(0.2,0.8,0.1,0.4)$ in the lower left panel and $(0.5,0.5,0.2,0.4)$ in the lower right panel, respectively. In all panels, the bounds labeled $1D$ are based on the combined rate $r_1+r_2$ and thus do not make use of the distinguishability of the two rates. The rates and scaled variances are given per unit time.}
    \label{fig:2d}
\end{figure}

\begin{figure}
    \centering
    % \tikzsetnextfilename{bound_heatmap_nn400_ls0p01}
    % \input{pdf/bound_heatmap_nn400_ls0p01.tikz}
    \includegraphics[scale=1]{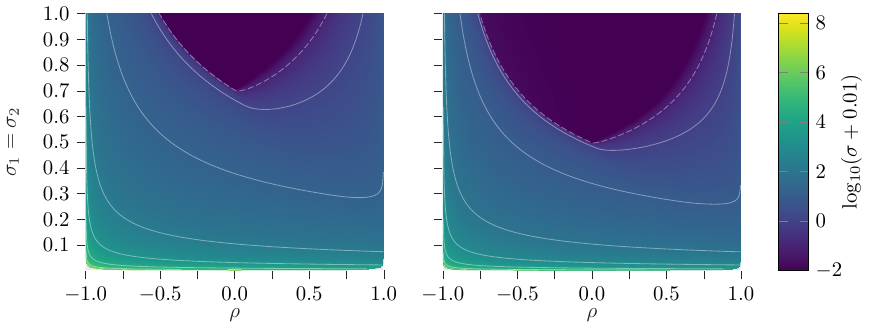}
    \caption{Bound on the steady-state entropy production rate $\sigma$ per unit time for two traffic observables (left) and for two flow observables (right). The bound \eqref{main} depends on five variables in the case with rates of two colors, of which we fix $\avg{r_1}=\avg{r_2}=0.5$ and set the two variances per unit time of the fluctuating rates equal, $\sigma_1^2 = \sigma_2^2$. In both cases, we vary $\sigma_1$ and the correlation $\rho\equiv\sigma_1\sigma_2/\Sigma_{12} = \sigma_1^2/\Sigma_{12}$ between $r_1$ and $r_2$. In the darkest regions where the entropy production vanishes, $\sigma=0$, the values of $\sigma_1,\sigma_2$ and $\rho$ can be realized in equilibrium. %{\color{orange}Note that the seamingly rounded corners for small $\sigma_1 = \sigma_2$ and $\rho$ near $\pm 1$ are an artifact of the used numerical method to determine $\sigma$ based on the bound \eqref{main} that becomes unstable.}
    }
    \label{fig:heatmap}
\end{figure}

In Figure \ref{fig:2d}, we illustrate the bounds \eqref{2d_flow} and \eqref{2d_traffic} for two flow observables and for two traffic observables, respectively, for different choices of mean rates $\avg{r_\alpha}$ and corresponding variances $\Sigma_{\alpha\alpha}\equiv\sigma_\alpha^2$ as a function of the correlation $\rho\equiv\sigma_1\sigma_2/\Sigma_{12}$ between the two rates. For comparison, we also display the bounds \eqref{1d_traffic} and \eqref{1d_flow}, labeled 1D, for which we combine the corresponding rates to $r\equiv r_1+ r_2$. Since the variance of this combined rate $r$ is given by $\Sigma_r=\sigma_1^2+\sigma_2^2+2\rho\sigma_1\sigma_2$, a positive correlation $\rho$ implies a lower precision of $r$ as compared to $\rho=0$ and thus, in general, a comparably smaller lower bound on the EPR. In contrast, anti-correlation of the two rates in $r$ requires a higher entropy production. The bounds \eqref{2d_flow} and \eqref{2d_traffic}, labeled 2D, are tighter than the 1D bounds based on the combined rate since they make use of the full information available. Mathematically, this stems from the convexity of the bounds \eqref{2d_flow} and \eqref{2d_traffic} as a function of the Fano factors $F_1$ and $F_2$, which are defined analogously to \eqref{fano}. In the upper left panel, both Fano factors are equal, so that the bounds labeled 2D coincide with their corresponding 1D bound for $\rho<\rho^*$ and deviate for $\rho\geq \rho^*$, where $\rho^*> 0$ must be determined numerically. In the other panels, the Fano factors of the two rates differ and thus using the full information in the 2D bounds yields an improvement over the corresponding 1D bounds for all correlations.

In Figure \ref{fig:heatmap}, we show the bounds \eqref{2d_flow} and \eqref{2d_traffic} as a function of variance and correlation of the rates with the two variances of the rates set equal. Since the bound based on the traffic is always tighter than the bound based on the flow, the parameter range of variances and correlations that can be realized in equilibrium is larger for the flow. Furthermore, for a high precision of the combined rate the required entropy production diverges due to either low standard deviations of the rates or strong anti-correlation between the rates.

%================================================================================================================================[ SECTION ]===========%
\section{Conclusion and Outlook}\label{sec:summary}

We have derived a model-independent lower bound on the entropy production rate of a Markov network based on averages and covariances of an arbitrary number of generalized fluctuating rates, starting from the multivariate Cram\'er-Rao inequality. Each of these generalized rates can be a time-antisymmetric current, a time-symmetric traffic or a time-asymmetric flow, which may be coarse-grained, i.e., consist of multiple indistinguishable microscopic rates of the same type. Furthermore, we have extended the bound to systems with time-dependent transition rates, thereby broadening its applicability to time-dependently driven systems and to relaxation processes. An overview of the relevant extant versions of the TUR and of the extensions and generalizations derived here is given in Table \ref{tab:comparison}.

The present framework is particularly suited for experiments where only coarse-grained counting statistics are accessible. Examples include measurements of molecular motors and enzymes or neural spike trains. In these settings, several microscopic transitions often contribute to the same experimentally observed event, while correlations between different counting observables can nevertheless be estimated directly from repeated trajectories. Our results, therefore, provide a practical route for inferring lower bounds on entropy production from experimentally accessible statistics without requiring the reconstruction of the underlying microscopic dynamics.

Throughout this work, we have assumed that mean values and covariances of the measured observables are known exactly. In realistic experiments, however, these quantities must themselves be estimated from finite data, introducing statistical uncertainty. Quantifying how estimation errors affect the estimator therefore constitutes a valuable next step. Moreover, since overdamped Langevin dynamics can be obtained as a suitable limit of Markov jump processes, we expect analogous bounds to hold for such a continuous stochastic dynamics. We expect such an extension to be relevant for experiments in which only low-dimensional projections of a high-dimensional system are observable.
\begingroup
\setlength{\tabcolsep}{7pt} % d 6pt
\renewcommand{\arraystretch}{1.5}
\begin{table}[H]
  \centering
  \caption{The TUR and its generalizations based on the precision of counting observables. The first conjecture or proof of each version of the TUR is referenced, with checkmarks representing our main results.\vspace{1.5\baselineskip}}
  \label{tab:comparison}
  \begin{tabular}{c|ccc|cccc}
    & \multicolumn{3}{c}{One color} & \multicolumn{4}{|c}{Multiple colors} \\ \cline{2-8}
    & Flow & Traffic & Current & Flows & Traffics & Currents & Combinations \\ \hline
    Nonequilibrium steady state & \cite{piet24} & \cite{piet24} & \cite{bara15,ging16} & \checkmark & \checkmark & \cite{dech18} & \checkmark \\
    Time-dependent processes & \checkmark & \checkmark & \cite{piet17,horo17,liu20,koyu20} & \checkmark & \checkmark & \checkmark & \checkmark \\ \hline
  \end{tabular}
\end{table}
\endgroup

\appendix
%================================================================================================================================[ SECTION ]===========%
\section{Complementary derivation of inequality \eqref{minimized} for traffics and flows using large deviation theory} \label{app:complderiv}
%================================================================================================================================[ SUBSECTION ]--------%
\subsection{Complementary entropy estimator based on large deviation theory}
\label{app:ld}
For a set of $\Omega$ observed rates consisting only of flow or traffic with one link per color, we can derive an alternative to bound \eqref{minimized} using large deviation theory for measurements in the long-time limit $T\to\infty$ as follows. The large-deviation rate function $I(\{\rho_i\},\{f_{ij}\})$ is given by \cite{seif25}
\begin{align}
    I(\{\rho_i\},\{f_{ij}\})= \sum_{ij} \rho_i k_{ij}-f_{ij}+f_{ij}\ln(f_{ij}/\rho_ik_{ij}).\label{rate_fn}
\end{align}
After contracting over the empirical densities $\rho_i$, i.e., choosing the most likely $\rho_i$, the rate function only depends on the empirical rates $f_{ij}$. Instead of performing the contraction, for which we would need to know the topology of a given network, an upper bound on the rate function can be found by using the non-optimal choice $\rho_i=p_i^s$, which results in
\begin{align}
    I(\{f_{ij}\})\leq \sum_{ij}\avg{f_{ij}}-f_{ij}+f_{ij}\ln(f_{ij}/\avg{f_{ij}})\equiv \sum_{ij}I_{ij}.\label{rate_fn_bound}
\end{align}
The two terms corresponding to the two directions of one link can further be bounded via \cite{ging16}
\begin{align}
    I_{ij}+I_{ji}\leq \frac{(\sigma_{ij} +\sigma_{ji} )(j_{ij}-\avg{j_{ij}})^2}{4\avg{j_{ij}}^2} \label{rate_fn_TUR}
\end{align}
where $\sigma_{ij} + \sigma_{ji}$ is the mean EPR at the link $ij$. We now use the bound \eqref{rate_fn_bound} for all observable links $\alpha\equiv ij$ and inequality \eqref{rate_fn_TUR} for all unobservable links. Furthermore, since inequality \eqref{rate_fn_bound} holds for any parameterization of the flows, we consider the case in which all observed rates are given by their true mean $\avg{r_\alpha} = \avg{r_{ij}}$ scaled with a factor, i.e., $r_{\alpha}=\nu_\alpha\avg{r_{\alpha}}$, and all currents are scaled by a uniform factor, $j_{ij}=\xi\avg{j_{ij}}$. With this parameterization and inequalities \eqref{rate_fn_bound} and \eqref{rate_fn_TUR}, an upper bound of the rate function $I(\{f_{ij}\}) = I(\xi,\{\nu_{\alpha}\})$ reads
\begin{align}
    I(\xi,\{\nu_{\alpha}\}) \leq & \left[\frac{\sigma}{2}-\sum_{\alpha} \avg{t_{\alpha}}  x_{\alpha}\artanh(x_{\alpha}) -\sum_{\beta} \avg{f_{\beta}}\frac{2x_{\beta}\artanh(x_{\beta})}{1+x_{\beta}}  \right]\frac{(\xi-1)^2}{2} \notag \\
    &+ \sum_{\alpha} \avg{t_{\alpha}} \left[1-\nu_{\alpha}+\frac{\nu_{\alpha}+\xi x_{\alpha}}{2}\ln\frac{\nu_{\alpha}+\xi x_{\alpha}}{1+x_{\alpha}}+\frac{\nu_{\alpha}-\xi x_{\alpha}}{2}\ln\frac{\nu_{\alpha}-\xi x_{\alpha}}{1-x_\alpha}\right] \notag \\
    &+ \sum_{\beta} \avg{f_{\beta}}\left[ 2(1 - \nu_\beta) - \frac{2 x_{\beta}(1-\xi)}{1+x_{\beta}}+\nu_{\beta}\ln \nu_{\beta}+\!\left(\nu_{\beta}-\frac{2\xi x_{\beta}}{1+x_{\beta}}\right)\ln\frac{\nu_{\beta}(1+x_{\beta})-2\xi x_{\beta}}{1-x_{\beta}}\right]. \label{ineq:xinu}
\end{align}
Next, we expand inequality \eqref{ineq:xinu} around the minimum of the right-hand side at $\xi=1$ and $\nu_\alpha=1$ up to second order in $\xi$ and $\{\nu_\alpha\}$ using the coefficients defined in \eqref{co2} and \eqref{co1} with currents set to zero. This expansion leads with $\boldsymbol{\delta} \equiv (\xi-1, \nu_1-1, \dots, \nu_\Omega-1)^\intercal$ to
\begin{align}
    I(\xi,\{\nu_{\alpha}\}) \leq  \max_{\mathbf{x}}\ \frac{\sigma}{2}\frac{(\xi-1)^2}{2}+\sum_\alpha \frac{\kappa_\alpha}{2}(\nu_\alpha-1)^2-\lambda_\alpha(\nu_\alpha-1)(\xi-1) -\frac{\rho_\alpha}{2}(\xi-1)^2 + O(\lVert\boldsymbol{\delta}\rVert^3).
\end{align}
Here, we have substituted $y_\alpha\equiv x_\alpha^2$.
Contracting over the unobservable ratio $\xi$ yields the rate function for the optimal $\xi-1 = \sum_\alpha \lambda_\alpha(\nu_\alpha-1)/(\frac{\sigma}{2}-\sum_\mu \rho_\mu)$,
\begin{align}
    I(\{\nu_{\alpha}\}) &\leq \max_{\mathbf{x}}\ \sum_\alpha \frac{\kappa_\alpha}{2}(\nu_\alpha-1)^2 -\frac{1}{2}\frac{\left[\sum_\alpha \lambda_\alpha(\nu_\alpha-1)\right]^2}{\frac{\sigma}{2}-\sum_\mu \rho_\mu}+ O(\nu_\alpha^3).\label{I}
\end{align}
Since the second derivatives of the rate function $I(\{\nu_{\alpha}\})$ are scaled entries of the inverse covariance matrix, $\partial \nu_\alpha\partial\nu_\beta I\big|_{\boldsymbol{\nu}=\boldsymbol{1}}=\Sigma^{-1}_{\alpha\beta}\avg{r_\alpha}\avg{r_\beta}/T$, we find
\begin{align}
    0 \preceq  \delta_{\alpha\beta}\kappa_\alpha-\frac{\lambda_\alpha\lambda_\beta}{\frac{\sigma}{2}-\sum_\mu \rho_\mu}-\frac{1}{T}\Sigma^{-1}_{\alpha\beta}\avg{r_\alpha} \avg{r_\beta}\equiv (M'_{\alpha\beta})\label{res_ld}
\end{align}
with the Kronecker delta $\delta_{\alpha\beta}$. This matrix inequality is a condition on whether a specific combination of mean rates, covariances and EPR $\sigma$ can be realized in any system. In other words, if $\boldsymbol{M}'$ is positive semi-definite for some $\mathbf{x}$, then this combination of $\avg{r_\alpha}, \Sigma^{-1}_{\alpha\beta}$ and $\sigma$ can be realized in a system. Thus, for measured covariances and means, the smallest value of $\sigma$ for which we can find an $\mathbf{x}$ such that $\boldsymbol{M}'\succeq 0$ is the smallest realizable value of $\sigma$ and thus a lower bound on the EPR of a given system.

The above derivation assumes that all generalized rates stem from one observed link each. However, the derivation also leads to the same result in the general case with an arbitrary number of links contributing to one color. In this general case, we combine the present derivation with the reasoning in Ref. \cite{piet24} in which the entropy estimator based on an observed traffic and the one based on an observed flow are equivalent to the bound \eqref{res_ld} specialized to a single observed color.

%================================================================================================================================[ SUBSECTION ]--------%
\subsection{Equivalence of the bounds \eqref{minimized} and \eqref{res_ld} without observed currents} \label{sec:equivalence}
Using $\chi\equiv \sigma/2 -\sum_\alpha \rho_\alpha$ in the case without observed currents, inequality \eqref{minimized} attains the simple form $\chi \geq \mathbf{\lambda}^\intercal \boldsymbol{M}^{-1}\mathbf{\lambda}$. The result obtained in \eqref{res_ld} based on large deviation theory demands instead that $\boldsymbol{M}' =  \boldsymbol{M}-\mathbf{\lambda}\mathbf{\lambda}^\intercal/\chi$ must be positive semi-definite. Since $\boldsymbol{M}$ is positive semi-definite, i.e., $M\succeq 0$, in both approaches, showing that the equivalence
\begin{align}
    \boldsymbol{M}-\frac{\mathbf{\lambda}\mathbf{\lambda}^\intercal}{\chi} \succeq 0 \qquad\iff\qquad \chi \geq \mathbf{\lambda}^\intercal \boldsymbol{M}^{-1} \mathbf{\lambda}
\end{align}
holds thus suffices to prove the equivalence of the inequalities \eqref{minimized} and \eqref{res_ld}.
By the Cauchy interlacing theorem for a rank-1 update \cite{horn13,leite22}, the term $-\mathbf{\lambda}\mathbf{\lambda}^\intercal/\chi$ can only change the sign of the smallest eigenvalue of $\boldsymbol{M}$. Thus, $\boldsymbol{M}'$ is positive definite if and only if its determinant is positive. Further, by the matrix determinant lemma, we have $\det\left(\boldsymbol{M}-\mathbf{\lambda}\mathbf{\lambda}^\intercal/\chi\right)=\det(\boldsymbol{M})\left(1-\mathbf{\lambda}^\intercal \boldsymbol{M}^{-1} \mathbf{\lambda}/\chi\right)$ and thus
\begin{align}
    \boldsymbol{M}-\frac{1}{\chi}\mathbf{\lambda}\mathbf{\lambda}^\intercal \succeq 0 \qquad\iff\qquad \det(\boldsymbol{M})\left(1-\frac{1}{\chi}\mathbf{\lambda}^\intercal \boldsymbol{M}^{-1} \mathbf{\lambda}\right) \geq 0 \qquad\iff\qquad \chi \geq \mathbf{\lambda}^\intercal \boldsymbol{M}^{-1} \mathbf{\lambda}.
\end{align}
Since $\chi \geq 0$ and $\boldsymbol{M}\succeq 0$, both derivations, and hence inequalities \eqref{minimized} and \eqref{res_ld}, are equivalent.

%================================================================================================================================[ SECTION ]===========%
\section{Saturation of the bounds \eqref{minimized} and \eqref{res_ld} without observed currents} \label{app:saturation}
A necessary condition for the saturation of the bound \eqref{res_ld} is that the network consists only of links that have either equal transition rates for both directions or are observable. For these networks, inequality \eqref{I} saturates being equivalent to the rate function \eqref{rate_fn} evaluated at its minimum, $I(\{p_i^s\},\{r_{\alpha}^s\})$ with steady-state proabilities $p_i^s$. Additionally, as a sufficient condition, the difference between the optimal $\rho_i^*$ (for which $\partial_{\rho_i} I(\{\rho_i^*\},\xi,\{\nu_\alpha\})=0$ with rate function as used in inequality \eqref{ineq:xinu}) and the stationary $p_i^s$ must be $\rho_i^* - p_i^s=O(\lVert\boldsymbol{\delta}\rVert^2)$ with $\boldsymbol{\delta}$ as defined in Appendix \ref{app:ld}, so that $I(\{p_i^*\},\xi,\{\nu_\alpha\}) = I(\{p_i^s\},\xi,\{\nu_\alpha\}) + O(\lVert\boldsymbol{\delta}\rVert^3)$ ensures the equivalence of \eqref{I} with the rate function \eqref{rate_fn} contracted over the empirical densities $\rho_i$.

We now provide two examples for which the bound \eqref{res_ld}, and thus \eqref{minimized} without observed currents, is saturated. First, in the special case of one color, saturation can be realized by a large unicyclic network, where one link is observed and all other links $ij$ have affinity $\ln(k_{ij}/k_{ji})\sim N^{-1}$, in the limit $N\to \infty$ of infinitely many links \cite{piet24}. More generally, any network in which a link is either observed or its affinity approaches zero in the limit of $N\to\infty$ links per cycle and that possesses transition rates leading to uniform $p_i^s$ and $\rho_i^*$ also saturates the bound \eqref{res_ld}.

\bibliographystyle{iopart-num}
\bibliography{references}

\end{document}